# *Machine Learning Driven Sensitivity Analysis of E3SM Land Model Parameters for Wetland Methane Emissions*

**Sandeep Chinta**
Center for Global Change Science, Massachusetts Institute of Technology, Cambridge, Massachusetts, USA.
sandeepc@mit.edu

**Xiang Gao**
Center for Global Change Science, Massachusetts Institute of Technology, Cambridge, Massachusetts, USA.
xgao304@mit.edu

**Qing Zhu**
Climate Sciences Department, Lawrence Berkeley National Laboratory, Berkeley, California, USA.
qzhu@lbl.gov

## Abstract

*Methane ($CH_4$) is the second most critical greenhouse gas after carbon dioxide, contributing to 16-25% of the observed atmospheric warming. Wetlands are the primary natural source of methane emissions globally. However, wetland methane emission estimates from biogeochemistry models contain considerable uncertainty. One of the main sources of this uncertainty arises from the numerous uncertain model parameters within various physical, biological, and chemical processes that influence methane production, oxidation, and transport. Sensitivity Analysis (SA) can help identify critical parameters for methane emission and achieve reduced biases and uncertainties in future projections. This study performs SA for 19 selected parameters responsible for critical biogeochemical processes in the methane module of the Energy Exascale Earth System Model (E3SM) land model (ELM). The impact of these parameters on various $CH_4$ fluxes is examined at 14 FLUXNET-$CH_4$ sites with diverse vegetation types. Given the extensive number of model simulations needed for global variance-based SA, we employ a machine learning (ML) algorithm to emulate the complex behavior of ELM methane biogeochemistry. ML enables the computational time to be shortened significantly from 6 CPU hours to 0.72 milliseconds, achieving reduced computational costs. We found that parameters linked to $CH_4$ production and diffusion generally present the highest sensitivities despite apparent seasonal variation. Comparing simulated emissions from perturbed parameter sets against FLUXNET-$CH_4$ observations revealed that better performances can be achieved at each site compared to the default parameter values. This presents a scope for further improving simulated emissions using parameter calibration with advanced optimization techniques like Bayesian optimization.*

# 1 INTRODUCTION

Methane ($CH_4$) is a potent greenhouse gas, responsible for approximately 20% of the warming potential as a result of anthropogenic activities since the start of the Industrial Revolution (Etminan et al., 2016). Although $CH_4$ is the second most influential greenhouse gas forcing global warming and climate change, following $CO_2$, its potency is further highlighted by the fact that the warming potential of $CH_4$ is estimated to be 28 times higher than that of $CO_2$ over a 100-year period, and 84 times higher over a 20-year period (Bridgham et al., 2013; IPCC, 2013). Atmospheric $CH_4$ concentrations have more than doubled since pre-industrial times, and this upward trend continues to persist (Dlugokencky et al., 2009; Jackson et al., 2020; Nisbet et al., 2019). The estimated annual growth rate of atmospheric $CH_4$ concentration for 2021 was a record high since 1984 (Lan et al., 2023) and more than three times the average annual growth rate from 2007 to 2015 (Poulter et al., 2017). Such an increase significantly contributes to the radiative forcing of the atmosphere and further amplifies global warming. Moreover, $CH_4$ has a large natural emission component from permafrost in the northern latitudes. Permafrost, once melting triggered by initial warming, leads to more emissions, followed by more warming and more emissions with a self-reinforcing cycle. Despite methane's relatively short atmospheric lifetime of 12.4 years (Balcombe et al., 2018), its warming potential makes it an essential cog in the wheel for measures to reduce climate change (Shindell et al., 2012).

While the impact of $CH_4$ on global warming is evident, it is essential to understand its sources to effectively manage and mitigate its release. $CH_4$ emissions originate from a broad spectrum of natural and anthropogenic sources, with marked variations observed in their relative contributions across various regions and timescales. Wetlands contribute to more than 30% of total emissions and are the most significant contributor to emissions among natural sources. The substantial contributors among anthropogenic sources are agriculture, fossil fuel extraction, and livestock farming (Bridgham et al., 2006; Ciais et al., 2013; Jackson et al., 2020; Kirschke et al., 2013; Saunois et al., 2016). Wetlands are diverse ecosystems consisting of swamps, marshes, and rice paddies, enabling $CH_4$ production through unique microbial metabolic processes within their anaerobic environments (Bodelier & Laanbroek, 2004; Turetsky et al., 2014). $CH_4$ emission from wetlands is challenging to measure and predict accurately due to their intricate nature and spatiotemporal variability (Rosentreter et al., 2021). In addition, the role of wetlands in the total $CH_4$ budget and their impact on inter-annual variability and changes in the $CH_4$ growth rate still needs to be fully understood (Poulter et al., 2017). This issue arises from various factors ranging from soil properties, temperature, vegetation types, and water table dynamics that control $CH_4$ production, consumption, and transfer in wetlands (Bousquet et al., 2006; Melton et al., 2013). Global warming could aggravate the $CH_4$ emissions from wetlands as they are susceptible to climatic conditions and land-use changes (Gurevitch & Mengersen, 2019). To address climate change effectively, it is critical that we enhance our ability to model and predict wetland $CH_4$ emissions. This requires comprehensive, process-based models that encompass all relevant factors and processes.

Biogeochemistry models inherently introduce uncertainties in modeling $CH_4$ emissions due to several factors. Model uncertainty that arises from each biogeochemistry model has its own simplifications (a combination of model structure, complexity, physics, usage, and tuning of model parameters) to represent real-world processes. Such simplifications can vary considerably among models and result in a wide range of fidelity. A large number of these model parameters relating to multiple physical, biological, and chemical processes associated with $CH_4$ dynamics induce parameter uncertainty. These parameters generally take fixed values, but they are not unambiguously known and usually must be prescribed based on the best available knowledge. Parameter uncertainty is commonly assessed by sensitivity analysis (SA) based on sampling within the theoretical, plausible ranges of parameters, which is the primary focus of this study (Müller et al., 2015; Ricciuto et al., 2021; Riley et al., 2011). Other sources of uncertainty include spatial variability of wetlands, scarcity of observations for calibration, initial and boundary conditions, and meteorological forcing to drive the model (Papa et al., 2013; Xu et al., 2012).

Sensitivity analysis quantifies the influence of different input parameters on the model's output, helping identify the parameters that contribute significantly to model parametric uncertainty. Performing global sensitivity analysis of biogeochemistry model parameters is critical to addressing the inherent parameter uncertainty. Several studies (Chinta et al., 2021; C. Wang et al., 2020) implemented sensitivity analysis in understanding parameter uncertainties in complex earth system models. Ricciuto et al. (2018) applied SA to the Energy Exascale Earth System Model (E3SM) land model (ELM) parameters with respect to carbon cycle output. Fisher et al. (2019) examined parameter controls on vegetation responses in the Community Land Model (CLM) using SA, and Yuan et al. (2021)examined the effects of warming and elevated $CO_2$ on peatland $CH_4$ emissions using a similar approach. Ricciuto et al. (2021) used sensitivity analysis and showed that production and substrate parameters are vital for regulating temporal patterns of surface $CH_4$ fluxes. Song et al. (2020) performed SA for a microbial functional group-based $CH_4$ model and observed that $CH_4$ emissions are sensitive to the parameters that regulate dissolved organic carbon and acetate production. However, a major challenge in sensitivity analysis is the efficient exploration of the parameter space. This involves a vast number of model simulations, making it computationally intensive. A full variance-based analysis typically needs thousands of model runs, like the Monte Carlo method, which is particularly demanding for complex biogeochemistry models. To address this, machine learning (ML)-based emulators, which mimic complex earth system models, have been introduced. These emulators present an efficient alternative, approximating model behavior accurately with fewer simulations. Müller et al. (2015) constructed an ML-based emulator for $CH_4$ parameter estimation in CLM4.5bgc. Gao et al. (2021) used emulators to quantify the sensitivity of soil moisture to uncertain CLM model parameters. Dagon et al. (2020) also implemented emulators in CLM biophysical parameter estimation.

Although SA and ML have been successfully employed in various areas of earth system modeling, their potential in improving $CH_4$ emissions modeling still needs to be explored. Our study addresses several primary research questions, such as: 1) which critical parameters dominate the sensitivity of model simulated $CH_4$ emission? 2) what are temporal (seasonal versus annual) and spatial (site-to-site or vegetation type to vegetation type) characteristics of such parametric sensitivity? and 3) is there any potential to improve model-simulated methane emissions? We integrated various advanced techniques to tackle these scientific questions. SA is employed to examine the influence of different input parameters on various components of $CH_4$ emissions, while ML is used to emulate the ELM biogeochemistry model with feasible computational cost desired by global SA. The paper is structured as follows: Section 2 presents the sensitivity analysis method and machine learning algorithm. Section 3 describes the model, FLUXNET-$CH_4$ sites, and the numerical experiment design. Section 4 presents the results and discussion. Section 5 concludes the paper and summarizes the key findings.

# 2 METHODS

## 2.1 Sobol Sensitivity Analysis

The Sobol sensitivity analysis method (Sobol, 2001), a variance-based approach to identify the sensitive model parameters, is used in this study. This method was successfully implemented in several studies (Baki et al., 2022a; Gao et al., 2021; Reddy et al., 2023; D. Ricciuto et al., 2018) to conduct sensitivity analysis for various Earth system model parameters. The Sobol method decomposes the total variance in the model output into the variances corresponding to either a single input parameter or a set of input parameters. There are two essential features of this method. First, it is a global method, as the sensitivity is evaluated across the whole input parameter space. Second, this method can quantify the primary or first-order effects of sensitivity for each parameter and the interaction effects between parameters. These features ensure a comprehensive understanding of the sensitivity analysis of the parameters is obtained.

The total output variance, $V$, is decomposed as

$$V = \sum_{i=1}^{n} V_i + \sum_{1 < i < j < n} V_{ij} + \ldots + V_{1,2,3,\ldots,n} \tag{1}$$

where $n$ is the total number of parameters, $V_i$ is the variance of $i^{th}$ parameter, $V_{ij}$ is the variance from the interactions of $i^{th}$ and $j^{th}$ parameters, and $V_{1,2,3,\ldots,n}$ is the variance from the interaction of all the $n$ parameters. As shown below, the Sobol indices are obtained by dividing the respective variances by the total variance.

$$S_i = \frac{V_i}{V}; \quad S_{ij} = \frac{V_{ij}}{V}; \quad \ldots; \quad S_{12\ldots n} = \frac{V_{12\ldots n}}{V} \tag{2}$$

where $S_i$ is the Sobol index for the first-order (main) effect from the $i^{th}$ parameter. Total order Sobol index of $i^{th}$ parameter, which is the sum of its main and all interaction effects, $S_{Ti}$ is given as:

$$S_{Ti} = S_i + S_{ij} + \ldots + S_{12..i..n} \tag{3}$$

Despite the detailed insights provided, this method requires multiple model runs to cover the entire parameter space for estimating the sensitivity indices accurately (Saltelli et al., 2008). The Monte Carlo approach is commonly employed to generate such a large number of parameter samples. Due to high computational demands, developing a machine learning-based emulator is essential and beneficial. Emulators use a significantly reduced number of model simulations to approximate the model behavior with little loss in accuracy. These emulators, once validated, can then be used to estimate the sensitivity indices.

## 2.2 Gaussian Process Regression

To address the computational demands of the Sobol method, Gaussian process regression (GPR), a machine learning algorithm, is employed to develop an emulator that approximates the model behavior. GPR is widely used as an emulator due to its robustness and flexibility (Rasmussen & Williams, 2006; Wang, 2020). This algorithm is particularly suitable when the relationship between inputs and output is complex and non-linear. Several studies (Baki et al., 2022b; Chinta & Balaji, 2020; Gong et al., 2015; C. Wang et al., 2014) established the superiority of GPR as an emulator for earth system models compared to other ML algorithms. GPR is defined by the mean function, $m(x)$, and covariance function, $k(x, x')$, where $x$ and $x'$ are points in the input space. The expected value of the function at point $x$ is given by the mean function, whereas the covariance function gives the correlation between the function values at two different points. For a Gaussian process $f(x) \sim GP(m(x), k(x, x'))$, the joint distribution of any finite number $(n)$ of function values $f = [f(x_1), f(x_2), \ldots, f(x_n)]^T$ follows a multivariate Gaussian distribution:

$$P(f|X) = \mathcal{N}(f|\mu, K) \tag{4}$$

where $X$ is the observations or training data, $\mu = [m(x_1), m(x_2), \ldots, m(x_n)]^T$ is the mean vector, and $K$ is the covariance kernel matrix with $K_{ij} = k(x_i, x_j)$.

The main advantage of GPR is the presence of a covariance function, which helps in encoding our assumptions about the function that is being learned. The mean function is usually chosen as a constant, with the value being either zero or the mean of the training data, which is also typically zero as the data is often normalized to a zero mean. GPR has several options to choose from for a covariance kernel function. Some commonly chosen ones are linear, constant, squared exponential, Matern kernel, and a combination of multiple kernels. One of the most widely used covariance kernel functions is the combination of constant kernel and radial basis function (RBF) kernel. This kernel function can be mathematically represented as:

$$k(x, x') = \sigma_f^2 \exp\left(-\frac{1}{2\ell^2} \|x - x'\|^2\right) \tag{5}$$

where $x$ and $x'$ represent two points in the input space, the two hyperparameters for this kernel are $\sigma_f^2$ (signal variance) and $l$ (length-scale). The signal variance controls the average distance of function values from their mean, while the length scale determines the smoothness of the function. This kernel function provides the GPR with the flexibility to capture complex patterns in the data. The hyperparameters of the kernel function can be learned from the training data using such techniques as maximum likelihood estimation. Once validated, the trained GPR emulator can not only predict the corresponding output for a new point in the input space but also quantify the degree of uncertainty in this prediction. This is a decisive advantage of GPR over other ML algorithms. The details of the experiment design for applying GPR in our study are presented in the next section.

## 3 EXPERIMENTAL FRAMEWORK

### 3.1 Model Description and Parameter Selection

The Energy Exascale Earth System Model (E3SM) land model version 2 (ELMv2) (Golaz et al., 2022) is used in this study. This model is branched from Community Land Model (CLM) version 4.5 (CLM4.5) (Oleson et al., 2013). The model underwent several updates since branching from CLM4.5 with a new biogeochemical representation of global carbon, nitrogen, and phosphorus cycles (Zhu et al., 2019). Some of the other updates include the introduction of the multiple agents' nutrient competition, dynamic allocation, a new photosynthesis physiology scheme, and new $N_2$ fixation and phosphatase modules. Several studies (Golaz et al., 2019, 2022;

Ricciuto et al., 2018) explained these updates from CLM4.5 in great detail. The $CH_4$ biogeochemistry model (Riley et al., 2011) simulates several processes, such as $CH_4$ production, ebullition, aerenchyma transport, aqueous and gaseous diffusion, $CH_4$ oxidation, and mass balance for unsaturated and saturated soils with the following governing diffusion equation:

$$\frac{\partial(RC)}{\partial t} = \frac{\partial F_D}{\partial z} + P(z,t) - E(z,t) - A(z,t) - O(z,t)$$

(6)

where $R$ represents gas in aqueous and gaseous phases, $C$ represents the concentration of $CH_4$ with respect to water volume (mol m$^{-3}$), $F_D$ represents aqueous and gaseous diffusion (mol m$^{-2}$ s$^{-1}$), $P$ represents $CH_4$ production (mol m$^{-3}$ s$^{-1}$), $E$ represents ebullition (mol m$^{-3}$ s$^{-1}$), $A$ represents aerenchyma transport (mol m$^{-3}$ s$^{-1}$), $O$ represents oxidation (mol m$^{-3}$ s$^{-1}$), z represents vertical dimension (m), and t represents time (s). Although the biogeochemistry model does not explicitly represent wetland plant functional types relevant to $CH_4$ production, the grid cell-averaged heterotrophic respiration rates are proxies for microbial substrate availability. These respiration rates are calculated using intrinsic turnover time for soil organic carbon, considering the impacts from environmental factors (e.g., temperature). $CH_4$ production rate is estimated after further accounting for $O_2$ limitation and being corrected for its soil temperature dependence, *pH*, the availability of electron acceptors associated with redox potential, and seasonal inundation fraction. Ebullition occurs when the $CH_4$ partial pressure, as a function of temperature and depth below the water table, exceeds 15% of the local pressure. Bubbles are added to the saturated columns' surface flux and placed immediately above the water table interface in unsaturated columns. Aerenchyma transport is modeled as gaseous diffusion driven by a concentration gradient between the specific soil layer and the atmosphere and, if specified, by vertical advection with the transpiration stream. $CH_4$ oxidation is represented with double Michaelis-Menten kinetics (Arah & Stephen, 1998), dependent on the gaseous $CH_4$ and $O_2$ concentrations. Gaseous diffusivity in soils depends on temperature-dependent molecular diffusivity, soil structure, porosity, and organic matter content. Aqueous diffusivity in the saturated part of the soil depends on temperature-dependent molecular diffusivity and porosity. Gaseous diffusion is assumed to dominate above the water table interface and aqueous diffusion below it.

These processes in the $CH_4$ biogeochemistry model are represented as functions of climate, vegetation, soil conditions, and empirical parameters. In the context of modeling wetland CH4 emission, uncertainty mainly comes from flux intensity and wetland extent. The global emission is the product of flux intensity and wetland extent. In this study, we only focus on the first part, assuming the wetland extent is 100% at site level, and use measured CH4 emission intensity to parameterize the model. The default values of these parameters are typically assigned based on the best available knowledge from a limited experimental or theoretical investigation. The parameters that influence methane emission, their default values and ranges are derived from Riley et al. (2011) and Koven et al. (2013). Table 1 presents the 19 ELM parameters used in this study, which pertain to various processes such as production, substrate availability, ebullition, diffusion, aerenchyma transport, and oxidation. For any parameter with an unknown uncertainty range, +/- 50% of the default value is used.

### 3.2 FLUXNET-CH4 data for wetland $CH_4$ emission

FLUXNET-$CH_4$ is a pioneering global network of sites that provides continuous, high-frequency, and quality-checked eddy covariance $CH_4$ flux measurements (Delwiche et al., 2021;

Knox et al., 2019). This data helps get a deeper understanding of the variability of $CH_4$ emissions worldwide and also help validate $CH_4$ emissions from biogeochemistry models. The network currently encompasses 81 sites across various vegetation types. From the initial 81 sites, we excluded crop sites due to the complexities introduced by irrigation management, such as quantifying the volume of water required for irrigation. Additionally, sites with less than two years of continuous observational data were omitted. Considering the computational expense of simulating all locations, we selectively chose our study sites to represent a diverse mix of vegetation types across various climate classifications.

Table 2 presents the list of 14 sites selected for this study, along with their locations and vegetation types. The vegetation types include needleaf evergreen temperate tree (NETT, 4 sites), broadleaf deciduous temperate tree (BDTT, 1 site), broadleaf deciduous boreal shrub (BDBS, 1 site), arctic c3 grass (AC3G, 1 site), cool c3 grass (CC3G, 6 sites), and warm c4 grass (WC3G, 1 site). The numerical PFT value represents the setting associated with the respective PFT in the ELM biogeochemistry model. The vegetation types are not distributed evenly across sites. Some vegetation types like NETT and CC3G are spread across multiple sites, while remaining vegetation types are attributed to only one site each.

### 3.3 Numerical Experiment Design

Site-level single-point ELM simulations are performed for the 14 FLUXNET-$CH_4$ sites. The methodology implemented in this study is presented in Fig. 1. A total of 190 simulations (10 times 19 selected parameters) (Loeppky et al., 2009) are performed for each site with different combinations of parameter values for each simulation. The 190 sets of different parameter values are generated using Latin Hypercube sampling (LHS) (McKay et al., 1979) across given parameter ranges (Table 1). LHS is a statistical method for efficiently generating numerous sets of parameter values from a multidimensional distribution. It is a type of stratified sampling that is superior to simple random sampling, especially for cases with a large number of dimensions. Each simulation follows the same 3-step modeling protocol. In the first step, an accelerated spin-up is performed for 300 years with $CO_2$ concentration set to the value of the year 1901. Climatic Research Unit and Japanese reanalysis (CRU JRA) v2.2 data (Harris , 2021) at a six-hourly frequency and 0.5° × 0.5° resolution is used for meteorological forcing. For each site, forcing data from the nearest grid point is used. The forcing data and $N_2$ depositions are cycled over the years 1901-1920. The second step involves a regular spin-up for 200 years with the same $CO_2$, $N_2$ deposition configuration and forcing data as in accelerated spin-up but without accelerating soil turnover. The third step is a 120-year transient run from 1901-2020. Time-varying historical $CO_2$ concentrations and CRU JRA forcing data and nitrogen depositions of the years 1901-2020 are used in this step. Each three-step simulation, spanning from the accelerated spin-up to the transient run, took 6 CPU hours to complete. Five model output variables are considered for sensitivity analysis: $CH_4$ emission, $CH_4$ production, diffusive surface $CH_4$ flux, ebullition surface $CH_4$ flux, and aerenchyma surface $CH_4$ flux. The values of these fluxes are averaged for 2001-2020.

**Table 1.** List of 19 ELM parameters used and their default values, ranges, and brief description.

| Mechanism | Parameter | Default | Range | Units | Description |
|---|---|---|---|---|---|
| Production | $Q_{10}$ | 2 | [1.5 4] | - | $CH_4$ production $Q_{10}$ |
| | $\beta$ | 0.2 | [0.1 0.3] | - | Effect of anoxia on decomposition rate |
| | $f_{CH_4}$ | 0.2 | [0.1 0.3] | - | Ratio between $CH_4$ and $CO_2$ production below the water table |
| Substrate availability | $z_\tau$ | 0.5 | [0.1 0.8] | m | e-folding depth for decomposition |
| | $\tau_{cwd}$ | 3.33 | [2 20] | year$^{-1}$ | Corrected fragmentation rate constant CWD |
| | $\tau_{l1}$ | 0.054 | [0.027 0.081] | year | Turnover time of litter 1 |
| | $\tau_{l2-l3}$ | 0.204 | [0.102 0.306] | year | Turnover time of litter 2 and litter 3 |
| | $\tau_{s1}$ | 0.137 | [0.0685 0.2055] | year | Turnover time of soil organic matter (SOM) 1 |
| | $\tau_{s2}$ | 5 | [0.0685 0.2055] | year | Turnover time of soil organic matter (SOM) 2 |
| | $\tau_{s3}$ | 222.22 | [111.11 333.33] | year | Turnover time of soil organic matter (SOM) 3 |
| Ebullition | $C_{e,max}$ | 0.15 | [0.075 0.225] | mol m$^{-3}$ | $CH_4$ concentration to start ebullition |
| Diffusion | $f_{D_0}$ | 1 | [1 10] | m$^2$ s$^{-1}$ | Diffusion coefficient multiplier |
| Aerenchyma | $p$ | 0.3 | [0.15 0.45] | - | Grass aerenchyma porosity |
| | $R$ | 2.9× 10$^{-3}$ | [1.45× 10$^{-3}$ 4.35× 10$^{-3}$] | m | Aerenchyma radius |
| | $r_L$ | 3 | [1.5 4.5] | - | Root length to depth ratio |
| | $F_a$ | 1 | [0.5 1.5] | - | Aerenchyma conductance multiplier |
| Oxidation | $K_{CH_4}$ | 5× 10$^{-3}$ | [5× 10$^{-4}$ 5× 10$^{-2}$] | mol m$^{-3}$ | $CH_4$ half-saturation oxidation coefficient (wetlands) |
| | $K_{O_2}$ | 2 × 10$^{-2}$ | [2× 10$^{-3}$ 2× 10$^{-1}$] | mol m$^{-3}$ | $O_2$ half-saturation oxidation coefficient |
| | $R_{o,max}$ | 1.25× 10$^{-5}$ | [1.25× 10$^{-6}$ 1.25× 10$^{-4}$] | mol m$^{-3}$ s$^{-1}$ | Maximum oxidation rate (wetlands) |

Table 2. Geographical and vegetation details of the simulated FLUXNET-$CH_4$ sites.

| Site ID | Site Name | Latitude | Longitude | PFT | PFT Name |
|---|---|---|---|---|---|
| RU-Fy2 | Fyodorovskoye dry spruce | 56.45 | 32.90 | 1 | Needleaf evergreen temperate tree |
| DE-SfN | Schechenfilz Nord | 47.81 | 11.33 | 1 | |
| CH-Dav | Davos | 46.82 | 9.86 | 1 | |
| US-Ho1 | Howland Forest (main tower) | 45.20 | -68.74 | 1 | |
| US-PFa | Park Falls/WLEF | 45.95 | -90.27 | 7 | Broadleaf deciduous temperate tree |
| RU-Cok | Chokurdakh | 70.83 | 147.49 | 11 | Broadleaf deciduous boreal shrub |
| SE-Deg | Degero | 64.18 | 19.56 | 12 | Arctic c3 grass |
| DE-Zrk | Zarnekow | 53.88 | 12.89 | 13 | Cool c3 grass |
| CH-Cha | Chamau | 47.21 | 8.41 | 13 | |
| DE-Hte | Huetelmoor | 54.21 | 12.18 | 13 | |
| US-OWC | Old Woman Creek | 41.38 | -82.51 | 13 | |
| US-WPT | Winous Point North Marsh | 41.46 | -83.00 | 13 | |
| CN-Hgu | Hongyuan | 32.85 | 102.59 | 13 | |
| US-MRM | Marsh Resource Meadowlands Mitigation Bank | 40.82 | -74.04 | 14 | Warm c4 grass |

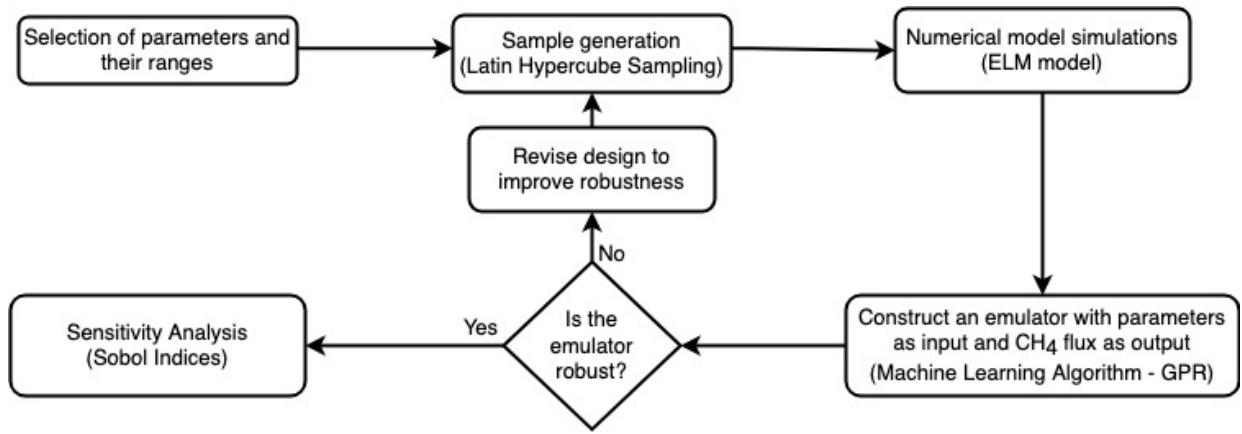

**Fig. 1:** Flowchart of the methodology implemented showing the main steps and sequence of operations.

### 3.4 Developing ML-based emulators using GPR

ML-based emulators were designed to take 19 parameter values as input and produce five CH4 flux values as outputs. For each of the five $CH_4$ fluxes, an individual emulator was developed at every site, resulting in a total of five emulators per site. Given the dependence of this study on emulator accuracy, we evaluated their performance using independent test data. Building emulator involves some assumptions and approximations. Unless the emulator correctly represents the simulator (model), Inferences made using that emulator will be invalid unless the emulator is able to correctly represent the simulator (model). To assess the adequacy of the emulator at untried points, additional 50 sets of LHS parameter values and model runs are generated at each site to validate the emulator. The coefficient of determination, $R^2$, between the model-simulated and emulator-estimated CH4 fluxes, served as our key evaluation metric. The closer to 1 the values are, the more accurate the emulator is. A single GPR prediction was achieved in just 0.72 milliseconds, in contrast to the 6 CPU hours required for the actual simulation, Once the emulator is validated, we then use 20-year averages of emulator-estimated monthly emissions to quantify two Sobol sensitivity indices (main and total effects) for five $CH_4$ fluxes in relation to each parameter (Sobol', 2003).

### 3.5 Comparing simulated emissions with FLUXNET-CH4 emissions

It is important to understand how the simulated emissions from perturbed parameter sets compare with observed emissions. The model simulated monthly-averaged $CH_4$ emissions from the 190 parameter sets are compared against FLUXNET-$CH_4$ observations at each site, and root mean square error (RMSE) is evaluated.

$$RMSE = \sqrt{\frac{\sum_{t=1}^{T}(sim^t - obs^t)^2}{T}} \quad (7)$$

where $sim^t$ and $obs^t$ are the simulated and observed values of monthly $CH_4$ emission from the simulated site at time $t$, respectively. $T$ is the number of months. A normalized root mean square error (nRMSE) was determined for the entire set of 190 initial sets of parameter values, with normalization performed based on the RMSE from the default run. The nRMSE for a given set, $i$, is computed as:

$$nRMSE_i = \frac{RMSE_i}{RMSE_{def}} \quad (8)$$

In this equation, $RMSE_i$ represents the RMSE for the specific set of parameter values, while $RMSE_{def}$ denotes the RMSE derived from the simulation with default parameter values. A parameter set with an nRMSE value less than 1 indicates improved performance (lower RMSE) in comparison to the default.

# 4 RESULTS AND DISCUSSION

## 4.1 Sensitivity Analysis - Main Effects and Interaction Effects

Emulators were developed for five CH4 fluxes at each site using the initial 190 simulations and then evaluated by comparing the emulator-predicted fluxes with the ELM-simulated counterparts from 50 independent test simulations at each site. The results for the CH-Cha site as an example (PFT-13: Cool c3 grass) are presented in Fig. 2. The emulators performed well for all the fluxes with $R^2$ values for test data ranging from 0.84 to 0.95. The emulators also performed reasonably well for fluxes at other sites with $R^2$ values above 0.80 (not shown). Overall, the emulator captures well the model behavior across the entire parameter uncertainty space for sites with different vegetation types and various fluxes and is considered to be accurate and robust. Therefore, the GPR-based emulators can be reliably applied to derive the Sobol SA indices and further quantify the main and interaction effects of the fluxes relative to each parameter.

Fig. 3 presents the heatmap of the main effect Sobol indices for various $CH_4$ fluxes concerning each model parameter. The main effect index of a parameter signifies the influence of that parameter alone on the flux, disregarding any interaction effects with other parameters. Heatmaps are exemplified for two sites, CH-Cha (PFT-13: Cool c3 grass) and SE-Deg (PFT-12: Arctic c3 grass). Each cell of the heatmap represents the value of the main effect index for its corresponding parameter (x-axis) and flux (y-axis), with the color intensity indicating the strength of the sensitivity. The heatmaps for both sites were remarkably similar, reflecting parallel sensitivity trends across the parameters at these two sites. The $CH_4$ production parameters $Q_{10}$ and $f_{CH_4}$ (ratio between $CH_4$ and $CO_2$ production below the water table) demonstrated pronounced sensitivity for all fluxes. The diffusion parameter $f_{D_0}$ (diffusion coefficient multiplier) was a predominantly sensitive parameter for diffusion, whereas the oxidation parameters $R_{o,max}$ (maximum oxidation rate - wetlands) and $K_{CH_4}$ ($CH_4$ half-saturation oxidation coefficient - wetlands) emerged as sensitive parameters for aerenchyma transport across these two sites. Apart from these five parameters, the remaining parameters had negligible influence on all fluxes for these two sites. Results corresponding to other sites are presented in Section 4.2.

Fig. 4 illustrates the total effect Sobol indices, encapsulating main and interaction effects, for various $CH_4$ fluxes relative to each parameter at the two sites mentioned above. This decomposition of the total effects of each parameter into main (blue) and interaction (red) effects is a vital attribute of the Sobol SA method that helps better understand the parameter sensitivity. The main effect values are the same as those represented in the heatmaps (Fig. 3). Main effects were more prominent than interaction effects for all parameters, highlighting the dominant influence individual parameters exert on different $CH_4$ fluxes. No parameter had a higher value of interaction effect than the main effect across all fluxes at these two sites. Other sites share the similar characteristics of total effects (not shown).

## 4.2 Sensitivity for Multiple Sites Across Vegetation Types

Fig. 5 illustrates the distribution of main effect sensitivity indices for each parameter across 14 FLUXNET-$CH_4$ sites corresponding to different $CH_4$ fluxes. These boxplots comprehensively represent the variation in sensitivity indices for multiple sites across vegetation types. Immediately evident is that production parameter $Q_{10}$ has the highest sensitivity (consistently high median values) among the parameters across all the fluxes, suggesting its significant role in various $CH_4$ fluxes across diverse geographical locations and vegetation types. Another production parameter $f_{CH_4}$ was also found to be a fairly sensitive parameter for all the fluxes. Contrary to the earlier heatmaps for two sites (Fig. 3) where $f_{D_0}$ did not influence emission, it was a sensitive parameter with the main effect value even higher than $Q_{10}$ for some sites. Diffusion flux was sensitive to diffusion parameter, $f_{D_0}$. Apart from these three parameters, other parameters like $R_{o,max}$, $K_{CH_4}$ were sensitive parameters for aerenchyma. Additionally, $z_\tau$ (e-folding depth for decomposition) was a sensitive parameter for some fluxes. Also, some parameters were sensitive at one or two sites, represented as outliers in the figure. It is important to note that roughly 13 parameters consistently showed minimal influence on various fluxes across all sites. These characteristics underlines the heterogeneity of $CH_4$ flux dynamics and associated parametric sensitivities across the sites (vegetation types). Examining the parametric sensitivities at more sites with the same vegetation type may help generalize their patterns.

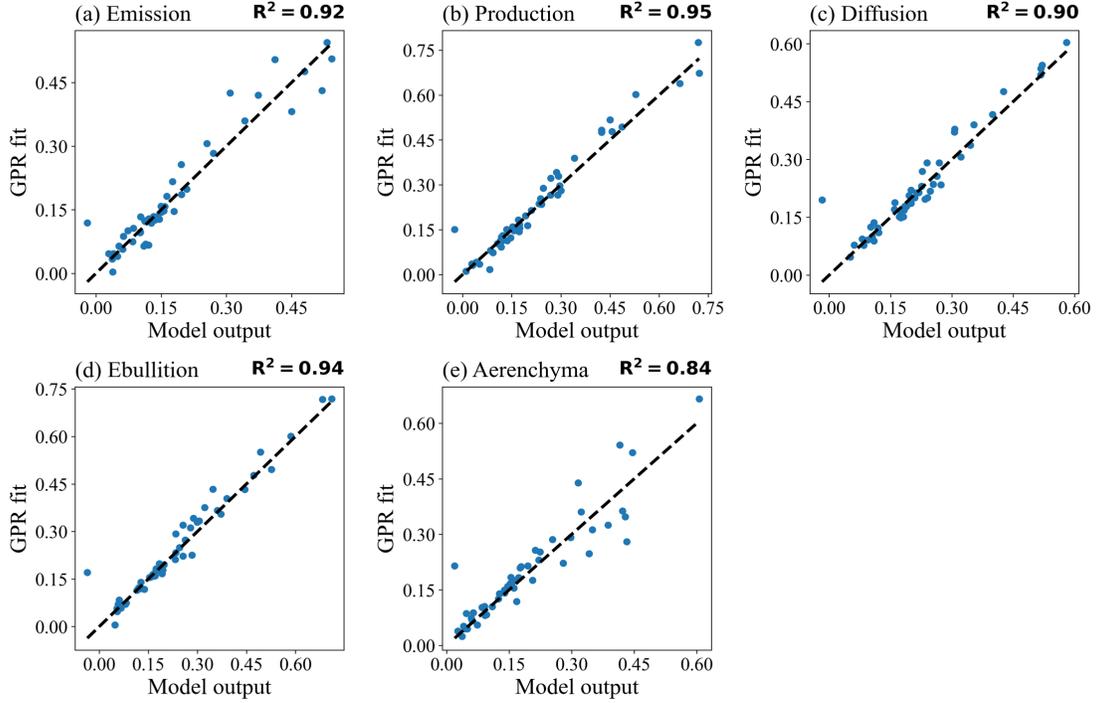

Fig. 2. Accuracy of the GPR model for test data (50 independent simulations) for different $CH_4$ fluxes at the CH-Cha site (PFT-13: Cool c3 grass). The horizontal axis denotes the model output, and the vertical axis represents the GPR fit.

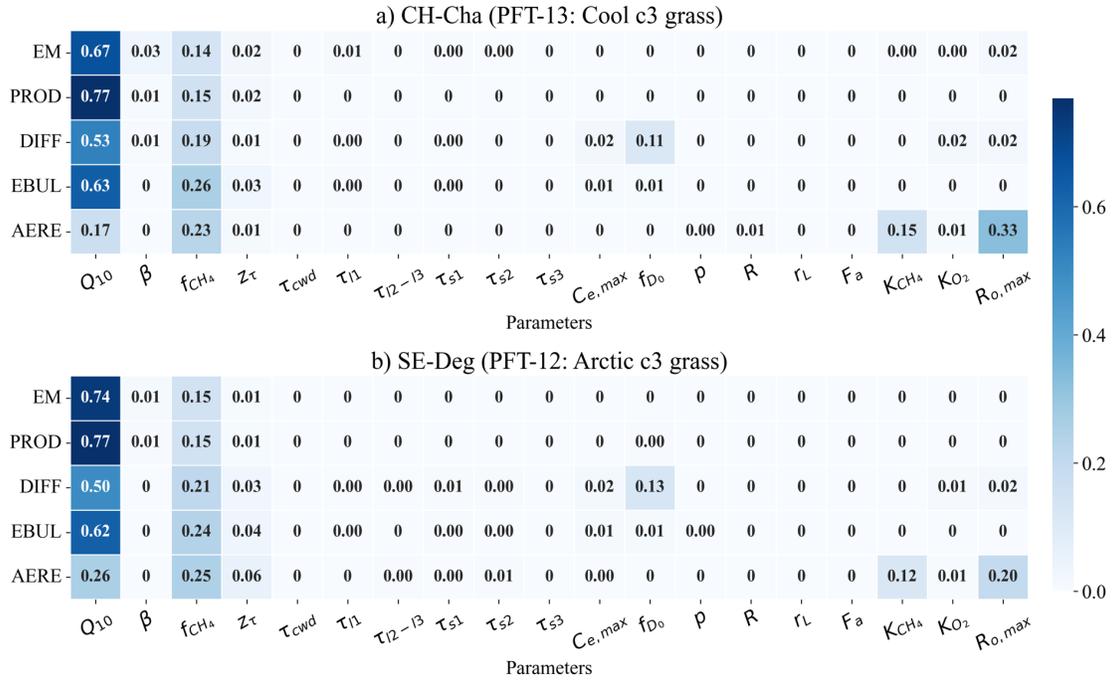

Fig. 3. Heat map of main effect sensitivity indices for different $CH_4$ fluxes (EM: Emission, PROD: Production, DIFF: Diffusion, EBUL: Ebullition, AERE: Aerenchyma) with respect to 19 parameters (shown in Table 1) for two sites a) CH-Cha and b) SE-Deg.

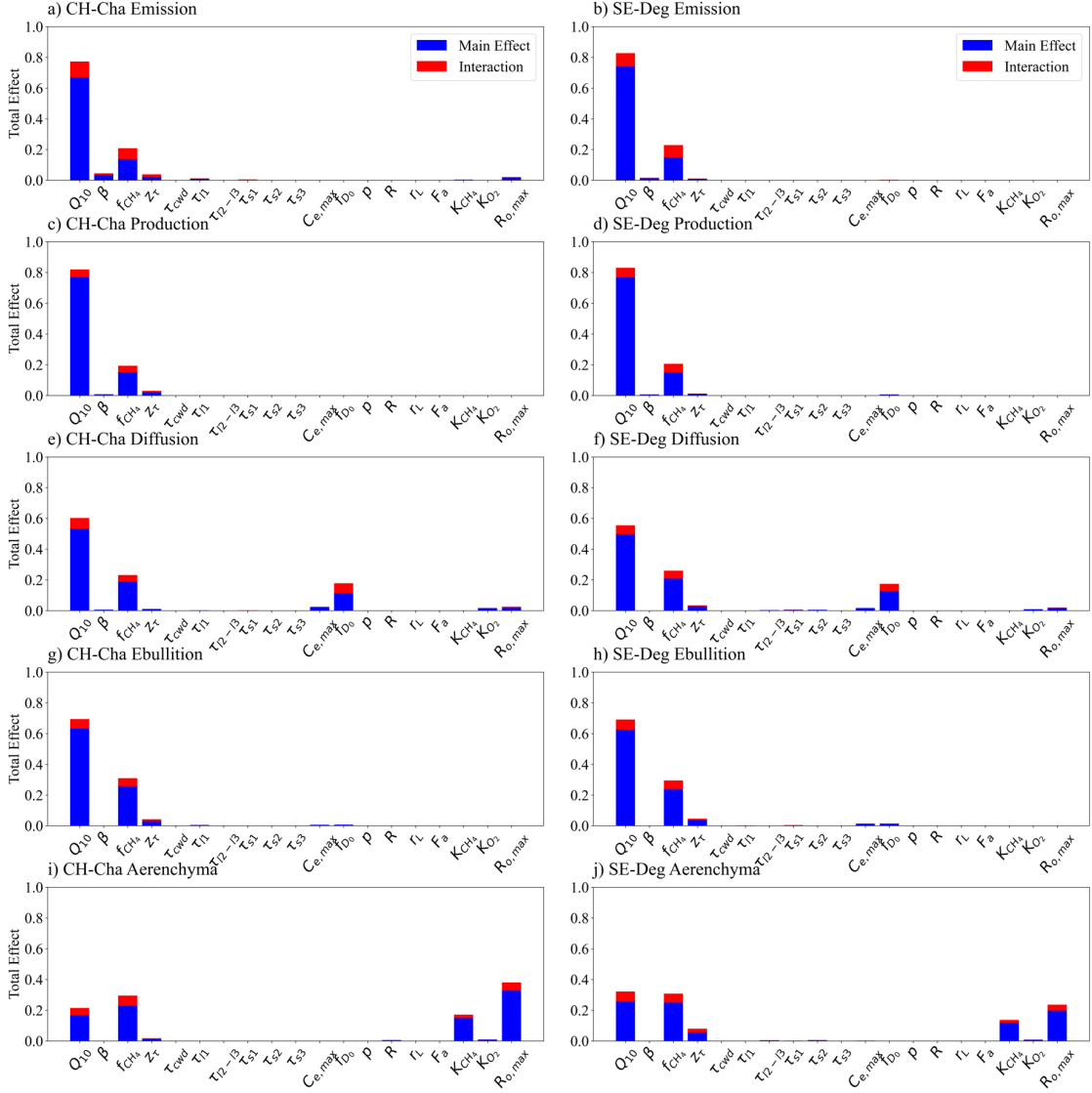

Fig. 4. The main effects and interaction effects (differences between the total and main effects) of 19 parameters for different $CH_4$ fluxes for two sites CH-Cha (PFT-13: Cool c3 grass) and SE-Deg (PFT-12: Arctic c3 grass).

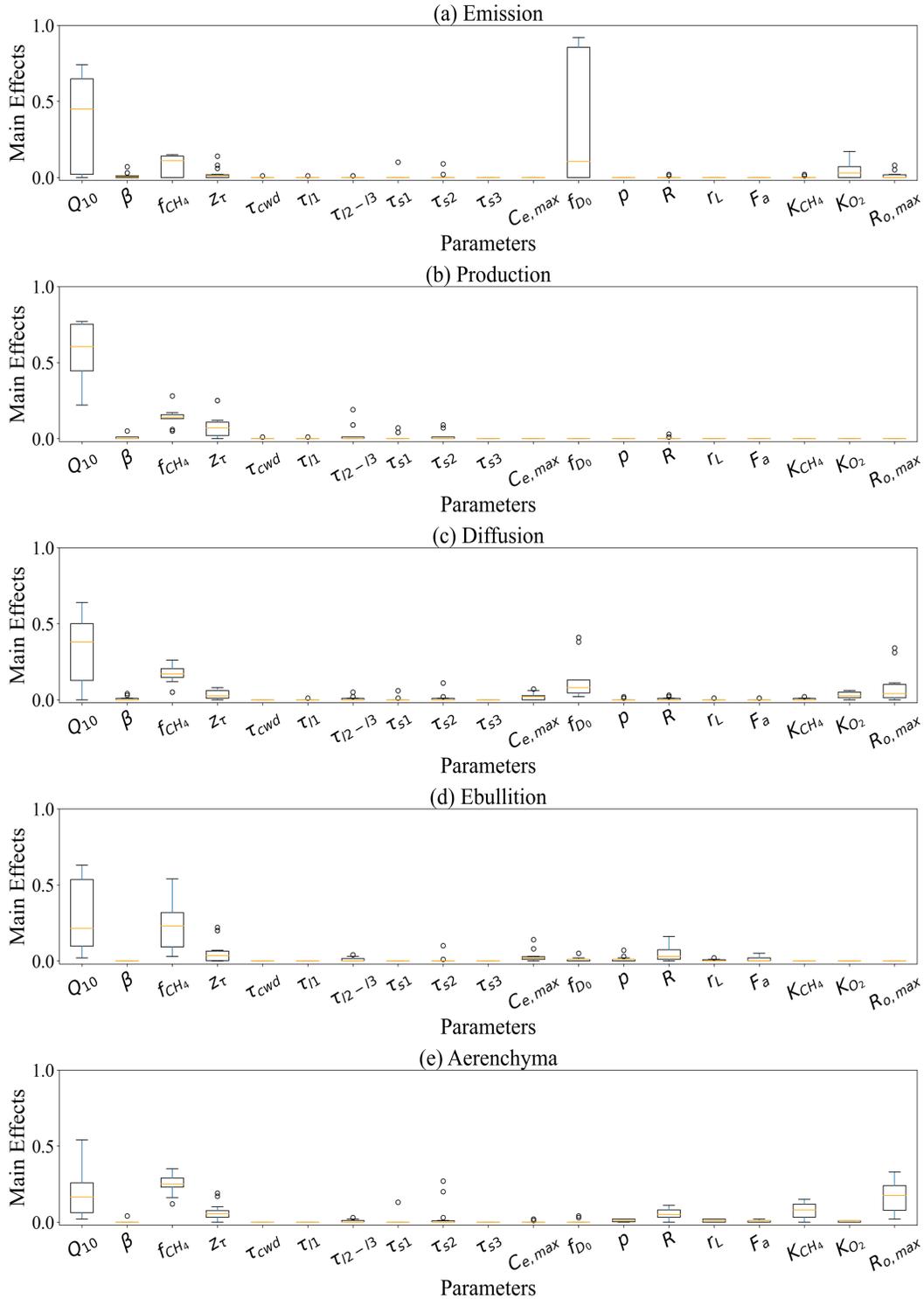

Fig. 5. Boxplots showing the distribution of main effect sensitivity indices for each parameter across 14 FLUXNET-CH$_4$ sites. The boxplot shows the median (orange line), interquartile range, minimum, and maximum after excluding outliers. An outlier, represented by a circle, is a data value outside 1.5 times the interquartile range above the upper quartile and below the lower quartile.

### 4.3 Contribution of Parameters to Variance in CH₄ Fluxes

Fig. 6 showcases a series of stacked bar plots representing the contribution of different parameters to the variance in $CH_4$ fluxes across the 14 FLUXNET-$CH_4$ sites, further categorized by their vegetation types. Each stacked bar corresponds to a specific site. The height of each parameter represents the percentage of the total effect index of that parameter with respect to the sum of the total effect indices of all parameters at that site, namely, the percentage of the total variance in $CH_4$ fluxes attributable to that parameter, including its interactions with other parameters. Only those parameters that contribute a minimum of 5% to the variance at any site have been included in the analysis. The relative size of a segment of a parameter indicates its proportional contribution to variance at that site.

The results aligned well with those shown in Fig. 5, with the production parameters $Q_{10}$ and $f_{CH_4}$, along with the diffusion parameter $f_{D_0}$ being the most influential parameters for different $CH_4$ fluxes across multiple sites. It was interesting to note that some sites had a combination of $Q_{10}$ and $f_{CH_4}$ as sensitive parameters for emission, whereas other sites had $f_{D_0}$ and $K_{O_2}$ as sensitive parameters. The production parameters $Q_{10}$ and $f_{CH_4}$ emerged as sensitive parameters for production at all sites. Apart from these two parameters, substrate availability parameters $z_\tau$, $\tau_{l2-l3}$ (turnover time of litter 2 and litter 3), and $\tau_{s2}$ (turnover time of soil organic matter 2) emerged as sensitive parameters for production at some sites. Substrate availability plays an important role in methane production at some sites as it determines the quantity and rate at which methanogenic microbes produce methane in anaerobic conditions. The diffusion parameter $f_{D_0}$ and the production parameters $Q_{10}$ and $f_{CH_4}$ emerged as sensitive parameters for diffusion at most of the sites. $R_{o,max}$ and $z_\tau$ emerged as sensitive parameters for diffusion at some sites. Ebullition was sensitive to the production parameters $Q_{10}$ and $f_{CH_4}$ at most sites. Ebullition parameter $C_{e,max}$ ($CH_4$ concentration to start ebullition) and aerenchyma parameters $R$ (aerenchyma radius) and $F_a$ (aerenchyma conductance multiplier) emerged as sensitive parameters for ebullition at some sites. The production parameters $Q_{10}$ and $f_{CH_4}$, oxidation parameters $R_{o,max}$, $K_{CH_4}$ and aerenchyma parameter $R$ were sensitive parameters for aerenchyma at most of the sites.

Across all sites, typically the top 5 or 6 parameters accounted for over 90% of the variance in $CH_4$ fluxes. However, in certain instances, even fewer parameters were responsible for a substantial portion of the variance. Notably, 13-14 parameters consistently showed a negligible effect on the $CH_4$ fluxes across these sites. For the four sites characterized by PFT-1 vegetation (Needleleaf evergreen temperate tree), the sensitive parameters remained largely consistent. This consistency prevailed irrespective of the individual climate classifications of these sites. This observation implies that, for Needleleaf evergreen temperate tree sites, the climate classification has minimal impact on parameter sensitivity. In contrast, the six sites with PFT-13 vegetation (Cool C3 grass) displayed more variability. While the first three of these sites shared a common sensitivity pattern, the next three differed in their sensitivities. This variation can be linked to their respective climate classifications as the first three sites are under a temperate climate, and the latter three are categorized as continental climate. This points to a stronger influence of climate classification on parameter sensitivity for Cool C3 grass sites.

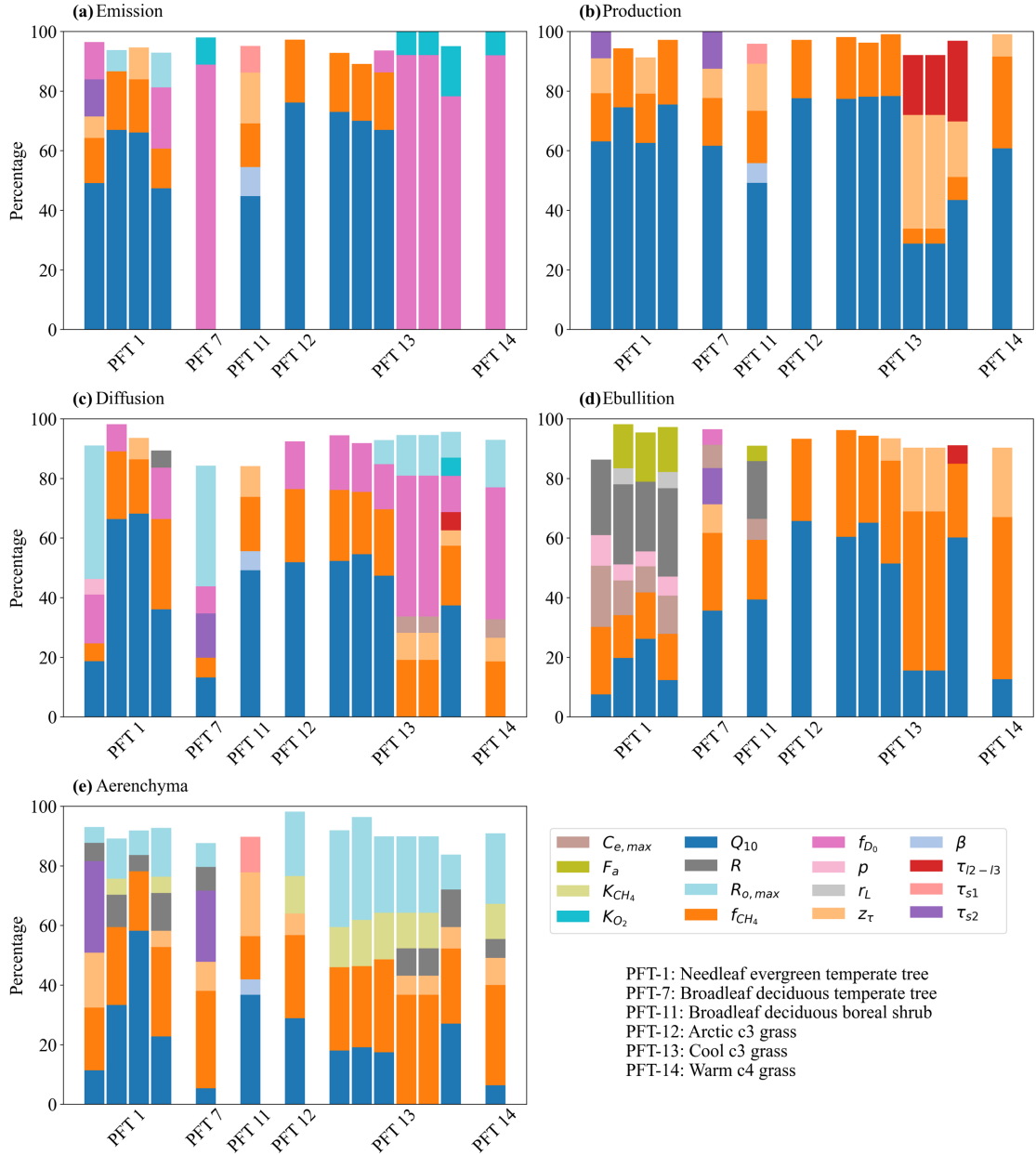

Fig. 6. Total effect sensitivity indices of parameters in the percentage of the variance in various $CH_4$ fluxes, grouped by vegetation types of 14 FLUXNET-$CH_4$ sites. Only those parameters with at least 5% contribution at any site are included.

In light of these findings, it is evident that while vegetation type plays a role in determining parameter sensitivity, climate classification can modulate this effect, especially for certain vegetation types. To arrive at a more definitive understanding, particularly for vegetation types with a single site, further analysis is necessary with a broader set of sites that share the same vegetation type.

## 4.4 Seasonal Characteristics of Parametric Sensitivity in CH$_4$ Emission Flux

Given the established seasonal variability in methane emissions from wetlands (Knox et al., 2021; Sakabe et al., 2021; Zhang et al., 2020), we sought to understand how the parametric sensitivity of methane emission fluctuates across months. For each of the five methane fluxes at every site, we constructed an emulator for each month using 20-year mean for that month. For instance, emissions for January were averaged from January 2001 through January 2020. Using these emulators, we evaluated monthly Sobol sensitivity indices. Fig. 7 shows the monthly variation in the main effect sensitivity indices of parameters for methane emission at 2 FLUXNET-CH$_4$ sites. Parameters with a minimum value of 0.05 for the main effect index for at least one month at the specific site were included in the analysis. The goal was to pinpoint sensitive parameters for each month concerning methane emission.

A distinct pattern was observed in the monthly sensitive parameters for the two sites, CH-Cha and SE-Deg. For CH-Cha (PFT-13: Cool c3 grass), the production parameters $Q_{10}$ and $f_{CH_4}$ were predominantly sensitive from December to March, whereas for SE-Deg (PFT-12: Arctic c3 grass), their sensitivity extended from November to June. In contrast, during the remaining months, the sensitivity was primarily associated with $f_{D_0}$ and $K_{O_2}$. The observed monthly fluctuations in parameter sensitivity can be linked to seasonal temperature variations, given that parameters $Q_{10}$ and $f_{D_0}$ are directly temperature-dependent (Riley et al., 2011). Other factors influencing the seasonal variation in methane emissions include gross primary productivity, ecosystem respiration, net ecosystem exchange, latent heat turbulent flux, soil temperature, water table depth, incoming shortwave radiation, and wind direction (Knox et al., 2021). This periodic parameter sensitivity behavior is distinctive from that from the 20-year annual mean in which neither $f_{D_0}$ nor $K_{O_2}$ were dominant sensitive parameters (Fig. 3). Long-term averages can sometimes mask specific temporal features by smoothing out variations from changing parameter values over shorter durations. Examining monthly averages reveals nuanced parametric sensitivity patterns that might be missed in long-term aggregates.

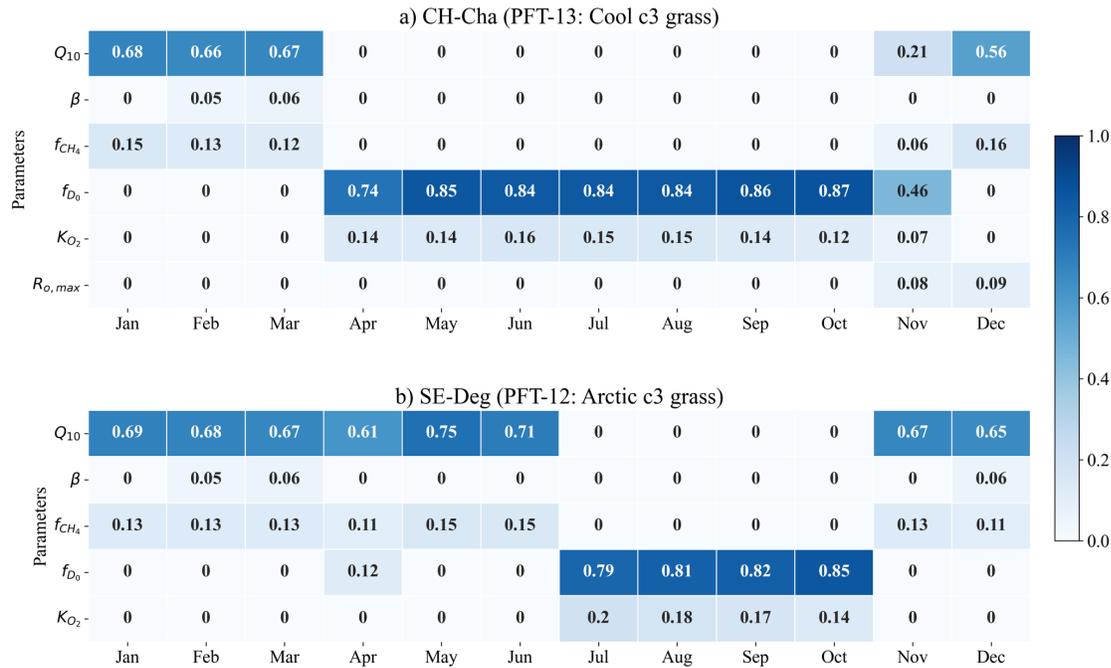

Fig. 7. Monthly fluctuation in main effect sensitivity indices of parameters of CH$_4$ emission for 2 FLUXNET-CH$_4$ sites. Only those parameters with a minimum value of 0.05 for the main effect index for at least one month at the specific site are included.

### 4.5 Parameter Ranking based on $CH_4$ Emission Flux Sensitivity

Fig. 8 presents the hierarchical ranking of parametric sensitivities to $CH_4$ emission across all sites. This ranking is derived based on the total effect sensitivity indices of parameters relative to emission obtained from the 20-year average. The total effects of parameters were averaged across all sites to offer a comprehensive perspective on their overall influence across diverse vegetation types. This averaging allows for capturing the general trends in parameter sensitivity and can help in identifying parameters of universal importance. Furthermore, the normalization of these averaged values ensures that the results are presented on a consistent scale of 0-100 to facilitate comparisons.

Five parameters $Q_{10}$, $f_{D_0}$, $f_{CH_4}$, $z_\tau$, and $K_{O_2}$ collectively accounted to approximately 95% of the normalized score. All the remaining parameters show little to negligible effect on $CH_4$ emission. $Q_{10}$ represents the temperature-dependent methane production. A higher $Q_{10}$ suggests an increase in temperature, increases methane production and emissions. Increased methane production due to a higher $Q_{10}$ can indirectly influence diffusion by creating larger concentration gradients to influence diffusion. The parameter $f_{CH_4}$ signifies the ratio between $CH_4$ and $CO_2$ production below the water table. A higher ratio signifies a greater dominance of methane in production and emission relative to $CO_2$. The diffusion coefficient multiplier, $f_{D_0}$, is equally important. This parameter directly alters the rate of methane movement through gas or liquid. A higher value of $f_{D_0}$ suggests more rapid methane diffusion. As the methane transport increases, it leads to higher emissions. The e-folding depth, $z_\tau$, determines the depth at which microbial decomposition diminishes exponentially. A greater e-folding depth suggests methane production can happen deeper, possibly causing a delay in its release or changing emission patterns due to its travel through various soil and water layers. The parameter $K_{O_2}$ indicates the oxygen concentration at which methane oxidation is halved. Higher $K_{O_2}$ values suggest that more methane is oxidized into $CO_2$, leading to reduced methane emissions.

The marginal influence of the remaining parameters on $CH_4$ emission suggests that while they may have site-specific importance, their overall contribution is subdued when averaged across all sites. This differentiation between universally influential parameters and those with localized effects can guide model developers to focus on the specific sensitive parameters for further improving $CH_4$ emission modeling based on different research objectives.

### 4.6 Comparison of Simulated Emissions with FLUXNET-$CH_4$ data

Sensitivity analysis was strictly a modeling exercise designed to understand how different parameters influence a model's output. Understanding how the simulated emissions from perturbed parameter sets compare with observed emissions is essential. This comparison works as an elementary assessment that allows us to understand whether there exists a potential to improve the simulated emissions with respect to observations by adjusting the parameter values within their ranges. We compared the model's simulated methane emissions from 190 parameter sets to FLUXNET-$CH_4$ observed emissions at each site. This comparison helps determine how the simulations from the 190 sets align with observations and how their performance stacks up against simulations using the default model parameters. Fig. 9 illustrates the variability in normalized root mean square error (nRMSE) values across 14 different FLUXNET-$CH_4$ sites, each grouped by their respective vegetation types. The box plots provide a comprehensive overview of the model performance of 190 parameter sets at each site. The RMSE from default parameters is represented by the red dashed line with an nRMSE value equal to 1. A parameter set with an nRMSE value less than 1 indicates an improved performance compared to the default. This is because the RMSE of that specific parameter set is lower than the RMSE obtained from the simulation with default parameter values.

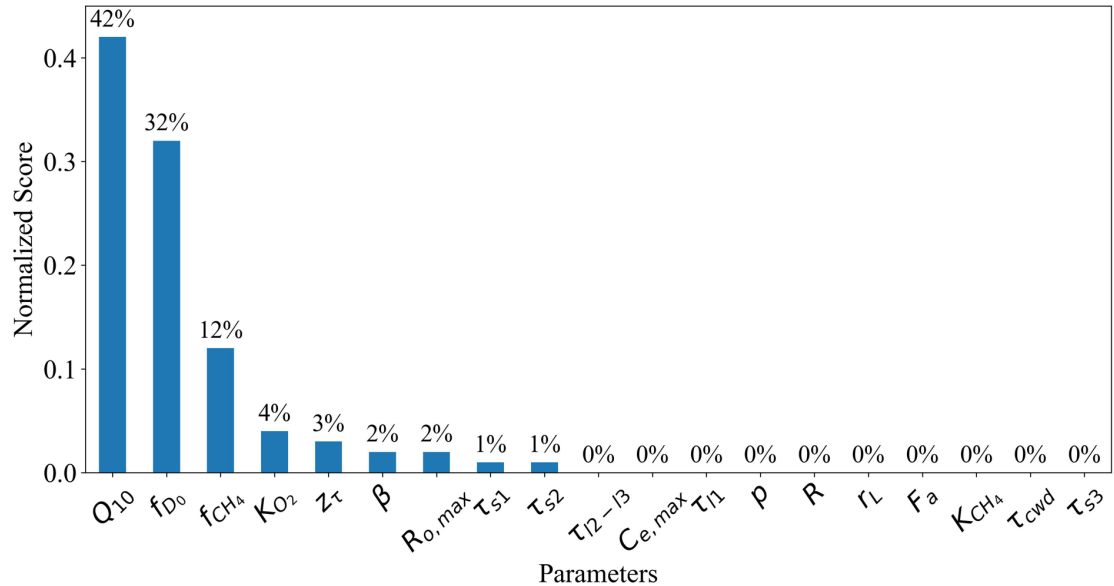

Fig. 8. Parameters ranked according to their sensitivity of annually averaged $CH_4$ emission across 14 FLUXNET-$CH_4$ sites. The percentage values over each bar represent the normalized score for that parameter.

A closer examination of the plot revealed significant variability in model performance across the sites. The median nRMSE value for some sites was higher than 1, whereas for other sites, it was lower than 1. For instance, sites like CH-Dav, US-PFa, and US-OWC were among the sites with higher median nRMSE values, and sites like DE-SfN, SE-Deg, and CH-Cha were among the sites with lower median nRMSE values. A higher median nRMSE value suggests potential challenges in accurately predicting methane dynamics for that site. On the other hand, sites like RU-Fy2, SE-Deg, and DE-Zrk had a tighter interquartile range, with their nRMSE values clustering closer, indicating a more consistent model performance for these locations even with perturbing parameter values. Furthermore, the presence of outliers in several sites highlights certain simulations with nRMSE values that significantly deviated from the majority. Outliers indicate that specific parameter combinations could either exceptionally enhance or hinder the model's performance at those sites.

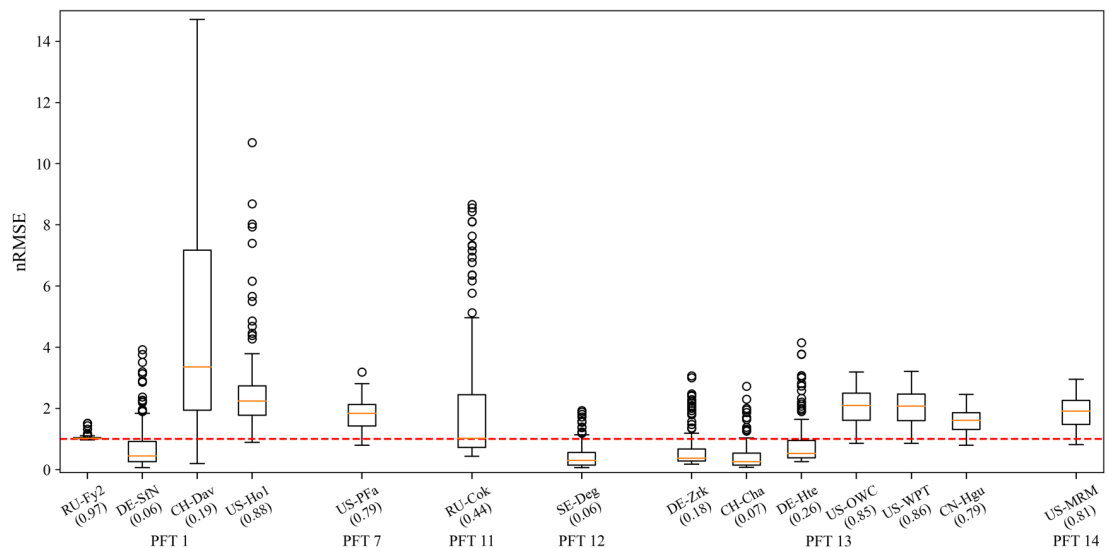

Fig. 9. Box plots depicting the normalized root mean square error (nRMSE) values across 14 FLUXNET-$CH_4$ sites, grouped by their vegetation types. The boxplot shows the median (orange line), interquartile range, minimum, and maximum after excluding outliers and individual circles marking outlier data points. The value in brackets below each site label denotes the minimum nRMSE value from a set of 190 simulations for that particular site. The red dashed line signifies an nRMSE value of 1, corresponding to the RMSE from the default parameter simulation for the respective site.

Notably, the minimum nRMSE values, provided in brackets for each site, underscore that there are alternative parameter simulations that can outperform the default for every site. This offers a promising avenue, particularly for model optimization tailored for each site. The values of the identified sensitive parameters can be adjusted within their respective ranges (Table 1) to minimize the difference between the model simulated and the observed $CH_4$ emissions at each FLUXNET-$CH_4$ site. This adjustment can be achieved systematically by employing an advanced optimization technique like Bayesian calibration (Gattiker et al., 2015; Kennedy & O'Hagan, 2001).

### 4.7 Limitations

This study offers significant insights into the sensitivity analysis of methane emissions. However, several inherent limitations need consideration. The presented results may be dependent on the choice of meteorological forcing data. The spatial scale discrepancy between the model simulations (0.5°) and observations (point-based) may also lead to some biases in the results. The emulator, despite its computational benefits, may not comprehensively represent the intricate and non-linear dynamics inherent in the ELM biogeochemistry model and this introduce slight discrepancies between the emulator's predictions and the actual ELM-simulated outputs. Additionally, using the Monte Carlo approach for generating large samples in the Sobol analysis introduces inaccuracies due to finite sample size (H. Wang et al., 2020). While the monthly fluctuations in parameter sensitivity were examined, the diurnal fluctuations were not explored. Notably, methane emissions exhibit significant diurnal variability (Knox et al., 2021), which could present another layer of complexity to the analysis. Future analyses could consider incorporating a broader range of sites spanning diverse vegetation types. This would ensure a more exhaustive assessment of parameter sensitivity across different ecosystems. Errors from external factors outside the methane biogeochemistry model, like heterotrophic respiration and net primary productivity, impact simulated methane emissions (Riley et al., 2011), which in turn affects the sensitivity analysis results. Despite these limitations, the findings from this study offer significant insights into the parametric sensitivity of various CH4 emissions.

## 5 CONCLUSIONS

This study carried out a sensitivity analysis of 19 E3SM model parameters with respect to methane emission from natural wetlands at 14 FLUXNET-$CH_4$ sites with diverse vegetation types. Machine learning-based emulators were employed to emulate the E3SM model in consideration of computational demands. The GPR-based emulators were shown to represent the model simulations reasonably well across all the sites. These emulators were used to calculate the Sobol sensitivity indices for various $CH_4$ fluxes. Five parameters $Q_{10}$ ($CH_4$ production), $f_{D_0}$ (diffusion coefficient multiplier), $f_{CH_4}$ (ratio between $CH_4$ and $CO_2$ production below the water table), $z_\tau$ (e-folding depth for decomposition), and $K_{O_2}$ ($O_2$ half-saturation oxidation coefficient) were identified as sensitive parameters across various fluxes and sites. These five sensitive parameters accounted for approximately 95% of the total variance for emission. Remarkably, around 14 parameters had negligible impact on emissions across all sites. Seasonal characteristics of parameter sensitivity to methane emissions showed specific features that long-term annual averages might overlook. Comparison of the model simulations against FLUXNET-$CH_4$ observations revealed a potential for improving simulated emissions via parameter calibration. Our future studies would focus on expanding this sensitivity analysis to more FLUXNET-$CH_4$ sites in order to better understand the dependence of parametric sensitivities on vegetation types and climatic conditions. The identified sensitive parameters can be systematically adjusted to reduce the simulation error with respect to observed methane emissions using Bayesian calibration and ML-based emulators. In addition, the availability of high-quality observations from a diverse range of wetlands will greatly benefit this exercise.

**Acknowledgments**

The authors thank Earle A Killian III and Waidy Lee MIT Seed Fund for their generous financial support of this research.